\documentclass[pra,amsmath,aps,10pt,superscriptaddress,letterpaper,tightenlines,showpacs]{revtex4}
\usepackage{bm}
\usepackage{stmaryrd}
\usepackage{amsmath}
\usepackage{amssymb}
\usepackage{graphicx}
\usepackage{textcomp}
\usepackage{calrsfs}
\usepackage{yfonts}
\usepackage{amsthm}
\usepackage{lipsum}
\newcommand{\p}{{\partial}}
\newcommand{\pt}{{\partial_t}}
\newcommand{\ptt}{{\partial^2_t}}
\newcommand{\curl}{{\nabla\times}}
\newcommand{\haf}{{\frac{1}{2}}}
\newcommand{\intf}{{\int_0^{\infty}\,}}
\newcommand{\la}{{\langle}}
\newcommand{\ra}{{\rangle}}

\begin{document}
%%%%%%%%%%%%%%%%%%%%%%%%%%%%%%%%%%%%%%%%%%%%%%%%%%%%%%%%%%%%%%%%%%%
\title{Electromagnetic field quantization in the presence of a rotating body}

\author{Fardin Kheirandish}
\email{fkheirandish@yahoo.com}
%\alternative{fardin_kh@phys.ui.ac.ir}
\author{Vahid Ameri}
\affiliation{Department of Physics,
Faculty of Science, The University of Isfahan,
Isfahan, Iran}
%%%%%%%%%%%%%%%%%%%%%%%%%%%%%%%%%%%%%%%%%%%%%%%%%%%%%%%%%%%%%%%%%%%
\begin{abstract}
Starting from a Lagrangian, the electromagnetic field is quantized in the presence of a body rotating along its axis of symmetry. Response functions and fluctuation-dissipation relations are obtained. A general formula for rotational friction and power radiated by a rotating dielectric body is obtained in terms of the dyadic Green's tensor. Hamiltonian is determined and possible generalizations are discussed. As an example, the rotational friction and power radiated by a spherical dielectric in the vicinity of a semi-infinite dielectric plane is obtained and discussed in some limiting cases.
\end{abstract}
\pacs{12.20.Ds, 42.50.Lc, 03.70.+k}
\maketitle
%%%%%%%%%%%%%%%%%%%%%%%%%%%%%%%%%%%%%%%%%%%%%%%%%%%%%%%%%%%%%%%%%%%
\section{Introduction}
%%%%%%%%%%%%%%%%%%%%%%%%%%%%%%%%%%%%%%%%%%%%%%%%%%%%%%%%%%%%%%%%%%%
A real physical system, whether classical or quantum, cannot be isolated from its surroundings. There are a wide range of physical problems described by a quantum field theory which has to be considered in the presence of a matter field. These matter fields are usually described by some bosonic or fermionic fields. For example, in quantum optics there are situations where the electromagnetic field should be quantized in the presence of a linear magnetodielectric medium \cite{Agarwal,Huttner,Vogel,Welsch,Matloob1,Stefano,Wubs1,Wubs2,Kheirandish1,Kheirandish2} or in calculating the static and dynamical Casimir forces \cite{Kheirandish3,Milton,Kardar,Emig,Neto,Milton2,Dalvit,Kennth,Jalal,Matloob2,Dodonov}. In these cases, the matter field should be included directly into the process of quantization in order to have a consistent formulation of the theory. There are two important situations corresponding to fixed and time-varying boundary conditions. For example electromagnetic field quantization in the presence of some static dielectrics is a problem with fixed boundary conditions leading to Casimir energy between fixed objects.  As an example of a time-varying boundary condition we can consider the electromagnetic field quantization in the presence of a moving dielectric \cite{Matloob3} or a scalar field constrained by Dirichlet boundary conditions on a moving boundary \cite{Fulling} or electromagnetic field in the presence of rotating objects \cite{Zeldovich,Rytov,Manjavacas,Maghrebi}. An important class of field theories with a time-varying boundary conditions are dynamical Casimir effects \cite{Dalvit}. Accelerated neutral objects are known to produce so called Casimir radiation by dynamical changes in the boundary conditions of the electromagnetic field associated to photon states \cite{Kennth, Kardar}.

The classical counterpart of this phenomenon is known as superradiance considered by Zel’dovich \cite{Zeldovich} where he
argues that a rotating object amplifies certain incident waves and when quantum mechanical considerations are applied, the object should spontaneously emit photons for some modes. Indeed this is shown to be the case for a rotating black hole by Unruh \cite{Unruh}.

The aim of the present paper is to investigate the electromagnetic field quantization in the presence of a dielectric body rotating along its axis of symmetry by generalizing the the ideas introduced in \cite{Huttner} in order to study quantum friction and amplifying ideas from a microscopic point of view beginning from a Lagrangian. For this purpose, we begin from a Lagrangian describing the whole system by modelling the medium with a continuum of harmonic oscillators \cite{Huttner} and follow a systematic approach based on canonical quantization of the whole system. In fact this is a generalization of the scheme introduced in \cite{Kheirandish1,Kheirandish2} for electromagnetic field quantization in the presence of a magnetodielectric medium (see Appendix for a brief review). General formulas for rotational friction and power radiated by a rotating body are obtained and in the case of small angular velocity or small radius of a dielectric sphere it is shown that the results coincide with those reported in \cite{Manjavacas}. As an example, the rotational friction and power radiated by a small spherical dielectric are obtained in the vicinity of a semi-infinite dielectric plane and limiting cases are discussed. Here for convenience we consider the matter field to be non-relativistic although generalization to a full covariant theory \cite{Amoo} and including magnetic properties is straightforward along the ideas introduced in \cite{Kheirandish1,Kheirandish2,Kheirandish3,Amoo} and may be dealt with in a separate work.
%%%%%%%%%%%%%%%%%%%%%%%%%%%%%%%%%%%%%%%%%%%%%%%%%%%%%%%%%%%%%%%%
\section{Lagrangian}
%%%%%%%%%%%%%%%%%%%%%%%%%%%%%%%%%%%%%%%%%%%%%%%%%%%%%%%%%%%%%%%%%%%
The coordinate-derivative transformations in rotating and fixed cylindrical reference systems are related by
\begin{eqnarray}\label{T}
&& \rho'=\rho,\,\,\,\varphi'=\varphi-\omega_0 t,\,\,\,z'=z,\,\,\,t'=t,\nonumber\\
&& \p_{\rho'}=\p_{\rho},\,\,\,\p_{\varphi'}=\p_{\varphi},\,\,\,\p_{z'}=\p_{z},\,\,\,\p_{t'}=\p_{t}+\omega_0\p_{\varphi},
\end{eqnarray}
here prime over coordinates denotes the rotating or body frame. A moving dipole produces magnetic moment which can interact with the magnetic component of the the electromagnetic field. Taking this point into account and using the transformations (\ref{T}) we can propose a Lagrangian in the fixed frame based on the Lagrangian structure in the body frame of the dielectric. Here we consider the non relativistic regime and for further convenience we assume that the dielectric is homogeneous in its rest frame. A full covariant theory including magnetic properties is also possible and may be dealt with in a separate work. Now let us consider the following Lagrangian for electromagnetic field in the presence of a dielectric rotating along its axis of symmetry namely $z$-axis with angular velocity $\omega_0$
\begin{eqnarray}\label{L}
\mathcal{L} &=& \haf\epsilon_0\,(\pt \mathbf{A})^2-\frac{1}{2\mu_0}(\curl\mathbf{A})^2+\haf\intf d\nu \,[(\pt \mathbf{X}+\omega_0\p_{\varphi}\mathbf{X})^2-\nu^2\mathbf{X}^2]\nonumber\\
&-& \epsilon_0\intf d\nu\,f_{ij}(\nu,t)X^j\pt A_i+\epsilon_0\intf d\nu\,f_{ij}(\nu,t)X^j (\mathbf{v}\times\curl\mathbf{A})_i.
\end{eqnarray}
The first term is the Lagrangian of the electromagnetic field in temporal gauge $(A^{0}=0)$. Note that in the Lab or fixed frame the electromagnetic field is a non rotating field and therefore, no modification is needed, the second term is the Lagrangian of the rotating dielectric which is modified trough the transformations (\ref{T}), the third and forth terms are the interaction between the dielectric and electromagnetic field inspired from electric $(-\mathbf{P}\cdot\mathbf{E})$ and magnetic $(-\mathbf{M}\cdot\mathbf{B})$ dipole interactions respectively. The local velocity $\mathbf{v}=\mathbf{\omega_0}\times\mathbf{r}$ is the velocity of a point with position $\mathbf{r}$ in the dielectric at time $t$ which is assumed to be non relativistic i.e $|\mathbf{v}|\ll c$. Note that in the third and forth terms the components of the dielectric field $X^j$ should be written in the fixed frame which is equivalent to considering a time-dependent coupling tensor $f_{ij}(\nu,t)$ defined by
\begin{equation}\label{C}
f_{ij} (\nu,t)=\left(
  \begin{array}{ccc}
    f_{xx} (\nu) \cos(\omega_0 t) & f_{xx} (\nu) \sin(\omega_0 t) & 0 \\
    -f_{yy} (\nu) \sin(\omega_0 t) & f_{yy} (\nu) \cos(\omega_0 t) & 0 \\
    0 & 0 & f_{zz}(\nu)\\
  \end{array}
\right).
\end{equation}
In the body frame $\omega_0=0$ and the coupling tensor is diagonal and also in this frame we assume $f_{xx} (\nu)=f_{yy} (\nu)$. Here we are considering homogeneous matter so the coupling tensor inside the matter is position independent and it is identically zero outside.
%%%%%%%%%%%%%%%%%%%%%%%%%%%%%%%%%%%%%%%%%%%%%%%%%%%%%%%%%%%%%%%%%%%
\section{Quantization and equations of motion}
%%%%%%%%%%%%%%%%%%%%%%%%%%%%%%%%%%%%%%%%%%%%%%%%%%%%%%%%%%%%%%%%%%%
From Lagrangian (\ref{L}) we find the corresponding conjugate momenta of the fields as
\begin{equation}\label{Pi}
\Pi_i (\mathbf{r},t)=\frac{\p \mathcal{L}}{\p (\pt A_i)}=-\epsilon_0 E_i-P_i=-D_i (\mathbf{r},t),
\end{equation}
\begin{equation}\label{Q}
Q_i=\frac{\p \mathcal{L}}{\p (\pt X_i)}=\pt X_i+\omega_0\,\p_{\varphi}X_i,
\end{equation}
where we have defined the polarization component $P_i (\mathbf{r},t)=\epsilon_0\intf d\nu f_{ij} (\nu,t)\,X^j (\mathbf{r},t,\nu)$ and $\mathbf{D}$ is the displacement vector. The system is quantized by imposing equal-time commutation relations
\begin{equation}\label{QA}
[A_i (\mathbf{r},t),\Pi_j (\mathbf{r}',t)]=i\hbar\,\delta_{ij}\,\delta (\mathbf{r}-\mathbf{r}'),
\end{equation}
\begin{equation}\label{QX}
[X_{i} (\mathbf{r},t,\nu),Q_{j} (\mathbf{r}',t,\nu')]=i\hbar\,\delta_{ij}\,\delta (\mathbf{r}-\mathbf{r}')
\delta (\nu-\nu').
\end{equation}
Now from Euler-Lagrange equations we find the equations of motion for the electromagnetic and matter fields respectively as
\begin{equation}\label{EA}
\frac{1}{c^2}\ptt\mathbf{A}+\curl\curl\mathbf{A} = \mu_0 [\pt\mathbf{P}-\curl (\mathbf{v}\times\mathbf{P})],
\end{equation}
and
\begin{eqnarray}\label{EX}
\ptt X_i+2\omega_0\,\p_{t}\p_{\varphi}\,X_i +\omega_0^2\p^2_{\varphi}X_i+\nu^2\,X_i &=& -\epsilon_0f_{ji}(\nu,t)\,[\pt\mathbf{A}-\mathbf{v}\times (\curl\mathbf{A})]_j,\nonumber\\
&=& -\epsilon_0f_{ji}(\nu,t)\,(\mathbb{D}_t \mathbf{A})_j,\\
\nonumber
\end{eqnarray}
where $\mathbb{D}_t=\pt-\mathbf{v}\times\curl\cdot$, is defined for notational convenience. The equation (\ref{EA}) in space-frequency is written as
\begin{equation}\label{FEA}
-\frac{\omega^2}{c^2}\,\mathbf{A}+\curl\curl\mathbf{A} = -i\omega\mu_0 [\mathbf{P}+\frac{1}{i\omega}\curl (\mathbf{v}\times\mathbf{P})]
=-i\omega\mu_0 \tilde{\mathbb{D}}\mathbf{P},
\end{equation}
where $\tilde{\mathbb{D}}=1+\frac{1}{i\omega}\curl (\mathbf{v}\times\cdot)$ \cite{Maghrebi}. By making use of the azimuthal symmetry we can expand the matter field components as
\begin{equation}\label{X-Expansion}
X_j (\mathbf{r},t,\nu)=\sum_{m} X_{j,m} (\rho,z,t,\nu) e^{i m\varphi}.
\end{equation}
Inserting (\ref{X-Expansion}) into (\ref{EX}) leads to
\begin{equation}\label{EXm}
\ptt X_{j,m}+2i m\omega_0\,\p_{t} X_{j,m} +(\nu^2-m^2\omega_0^2) X_{j,m}=-\epsilon_0f_{ij}(\nu,t)\,(\mathbb{D}_t \mathbf{A})_{i,m},
\end{equation}
with the formal solution
\begin{equation}\label{SX}
X_{j,m} (\rho,z,t,\nu)=X^{N}_{j,m} (\rho,z,t,\nu)-\epsilon_0\int dt'\,G(t-t',m,\nu)\,f_{ij} (\nu,t')\,(\mathbb{D}_{t'} \mathbf{A})_{i,m},
\end{equation}
where the Green's function is given by
\begin{equation}\label{Green-0}
G(t-t';m,\nu)=e^{i m\omega_0 (t-t')}\,\frac{\sin[(\nu(t-t')]}{\nu}\,\theta(t-t'),
\end{equation}
and the homogeneous solution $X^{N}_{i,m}$, interpreted as a noise field, is
\begin{equation}\label{HOMO}
X^{N}_{i,m} (\rho,z,t,\nu)=a^{\dag}_{i,m}(\rho,z,\nu)e^{i(\nu-m\omega_0)t}+a_{i,-m}(\rho,z,\nu)e^{-i(\nu+m\omega_0)t}.
\end{equation}
Now from (\ref{X-Expansion}) the noise field component $X^{N}_i$ can be expanded in terms of the ladder operators in the body frame as
\begin{eqnarray}\label{NField}
X^{N}_i (\rho,\varphi,z,\nu,t)=\sum_{m}\bigl[e^{i m\varphi} e^{i (\nu-m\omega_0)t}\,a^{\dag}_{i,m} (\rho,z,\nu)+e^{-i m\varphi}e^{-i (\nu-m\omega_0)t}\,a_{i,m} (\rho,z,\nu)\bigr],
\end{eqnarray}
and from canonical quantization rules (\ref{QX}) we find
\begin{equation}\label{amamdag}
[a_{i,m} (\rho,z,\nu),a^{\dag}_{j,m'} (\rho',z',\nu')] =\frac{\hbar}{4\pi\nu}\,\delta_{ij}\,\delta_{mm'}\delta(\nu-\nu')\,\frac{\delta(\rho-\rho')
\delta(z-z')}{\rho}.
\end{equation}
The Hamiltonian corresponding to to the noise field $X^{N}$ in the body frame is the thermal bath defined by
\begin{eqnarray}
H_B &=& \haf\intf d\nu\int d\mathbf{r}\,[(\pt \mathbf{X}_\nu)^2+\nu^2 \mathbf{X}^2_\nu],\nonumber \\
   &=& \sum_{n,j}\intf d\nu\int \rho d\rho dz\, \nu^2 (\hat{a}^\dag_{j,n}\hat{a}_{j,n}+\hat{a}_{j,n}\hat{a}^\dag_{j,n}),\nonumber \\
   &=&  \sum_{n,j}\intf d\nu\int \rho d\rho dz\, \frac{\hbar\nu}{2} (\hat{b}^\dag_{j,n}\hat{b}_{j,n}+\hat{b}_{j,n}\hat{b}^\dag_{j,n}),
\end{eqnarray}
where in the last equality we have defined the normalized ladder operators $\hat{b}_{j,n}(\hat{b}^\dag_{j,n})=\sqrt{\frac{2\nu}{\hbar}}\hat{a}_{j,n}(\hat{a}^\dag_{j,n})$ to resemble a continuum of independent quantum harmonic oscillators. From equilibrium quantum statistical mechanics we have
\begin{eqnarray}
  \la \hat{b}^\dag_{i,m}\hat{b}_{j,n} \ra &=& tr[\hat{\rho}\, \hat{b}^\dag_{i,m}\hat{b}_{j,n}]=\delta_{ij}\delta_{mn}\delta(\nu-\nu')
  \frac{\delta(\rho-\rho')\delta(z-z')}{2\pi\rho}\frac{1}{e^{\frac{\hbar\nu}{k T}}-1},\nonumber\\
  \la \hat{b}_{i,m}\hat{b}^\dag_{j,n} \ra &=& tr[\hat{\rho}\, \hat{b}_{i,m}\hat{b}^\dag_{j,n}]=\delta_{ij}\delta_{mn}\delta(\nu-\nu')
  \frac{\delta(\rho-\rho')\delta(z-z')}{2\pi\rho}\frac{e^{\frac{\hbar\nu}{k T}}}{e^{\frac{\hbar\nu}{k T}}-1}.\nonumber\\
\end{eqnarray}
Therefore, if the matter field or dielectric body is held in temperature $T$ then
\begin{equation}\label{FLUC-a}
\la a^{\dag}_{i,m} (\rho,z,\nu)\,a_{j,m'} (\rho',z',\nu')\ra_T=\frac{\hbar}{2\nu}\,n_{T} (\nu)\,\delta_{mm'}\delta_{ij}\,\delta(\nu-\nu')\,\frac{\delta(\rho-\rho')\delta(z-z')}{2\pi\rho},
\end{equation}
where $n_{T}(\omega)=[\exp(\hbar\omega/k T)-1]^{-1}$ is the thermal mean number of photons and we have switched to the non-normalized operators. From (\ref{SX}) and definition of the polarization we have
\begin{equation}\label{PO}
P_{k,m} (\rho,z,t)=P^N_{k,m} (\rho,z,t)-\epsilon_0^2\int dt'\intf d\nu\,f_{kj} (\nu,t) f_{ij} (\nu,t')\,G(t-t';m,\nu)\,
(\mathbb{D}_{t'} \mathbf{A})_{i,m},
\end{equation}
where $P^N_{k,m}$ are the fluctuating or noise polarization components defined by
\begin{equation}\label{N}
P^N_k (\mathbf{r},t)=\epsilon_0\intf d\nu\,f_{ki}(\nu)\,X^N_i (\mathbf{r},\nu,t).
\end{equation}
If we define the following response function
\begin{equation}\label{RES}
\chi^{ee}_{kj}(t-t',m) = \epsilon_0\intf d\nu\, G(t-t';m,\nu)\,f_{kl} (\nu,t) f_{jl} (\nu,t'),
\end{equation}
then using (\ref{C}) we can easily find
\begin{eqnarray}\label{RES1}
\chi^{ee}_{zz}(t-t',m) &=& \epsilon_0\intf d\nu\, G(t-t';m,\nu)\,f^{2}_{zz} (\nu),\nonumber\\
\chi^{ee}_{xx}(t-t',m)=\chi^{ee}_{yy}(t-t',m) &=& \epsilon_0\intf d\nu\, G(t-t';m,\nu)\,f^{2}_{xx} (\nu) \cos[\omega_0 (t-t')],\nonumber\\
\chi^{ee}_{xy}(t-t',m)=-\chi^{ee}_{yx}(t-t') &=& \epsilon_0\intf d\nu\, G(t-t';m,\nu)\,f^{2}_{xx} (\nu) \sin[\omega_0 (t-t')],\\
\nonumber
\end{eqnarray}
with the following Fourier transforms
\begin{eqnarray}\label{RES2}
\chi^{ee}_{zz}(\omega,m) &=& \epsilon_0\intf d\nu\, \frac{f^{2}_{zz} (\nu)}{\nu^2-(\omega-m\omega_0)^2},\nonumber\\
\chi^{ee}_{xx}(\omega,m)=\chi^{ee}_{yy}(\omega,m) &=& \epsilon_0\intf d\nu\, \frac{1}{2}\bigg[\frac{f^{2}_{xx} (\nu)}{\nu^2-(\omega_{+}-m\omega_0)^2}+\frac{f^{2}_{zz} (\nu)}{\nu^2-(\omega_{-}-m\omega_0)^2}\bigg],\nonumber\\
\chi^{ee}_{xy}(\omega,m)=-\chi^{ee}_{yx}(\omega,m) &=& \epsilon_0\intf d\nu\, \frac{1}{2i}\bigg[\frac{f^{2}_{xx} (\nu)}{\nu^2-(\omega_{+}-m\omega_0)^2}-\frac{f^{2}_{zz} (\nu)}{\nu^2-(\omega_{-}-m\omega_0)^2}\bigg].\\
\nonumber
\end{eqnarray}
where $\omega_{\pm}=\omega\pm\omega_0$. One can easily show that the response functions in the body frame denoted by $\chi^{0}_{kj}(\omega)$ are given by
\begin{equation}\label{kapa-0}
\chi^{0}_{kk}(\omega) =\epsilon_0\intf d\nu\, \frac{f^{2}_{kk} (\nu)}{\nu^2-\omega^2},
\end{equation}
and this recent relation leads to
\begin{equation}\label{kapa-0}
\frac{f^2_{kk} (\nu)}{\nu}=\frac{2}{\pi\epsilon_0}\,\mbox{Im}[\chi^0_{kk} (\nu)].
\end{equation}
Now using (\ref{RES2}) one can find the relations between the response functions in body and Lab frames as
\begin{eqnarray}\label{connection}
&& \chi^{ee}_{zz}(\omega,m)=\chi^{0}_{zz}(\omega-m\omega_0),\nonumber\\
&& \chi^{ee}_{xx}(\omega,m)=\chi^{ee}_{yy}(\omega,m)=
\frac{1}{2}[\chi^{0}_{xx}(\omega_{+}-m\omega_0)+\chi^{0}_{xx}(\omega_{-}-m\omega_0)],\nonumber\\
&& \chi^{ee}_{xy}(\omega,m)=-\chi^{ee}_{yx}(\omega,m)=
\frac{1}{2i}[\chi^{0}_{xx}(\omega_{+}-m\omega_0)-\chi^{0}_{xx}(\omega_{-}-m\omega_0)].\\
\nonumber
\end{eqnarray}
Also from (\ref{PO},\ref{RES2}) we find
\begin{eqnarray}\label{PO-compact}
&& P_{z} (\rho,\varphi,z,\omega)=P^{N}_{z} (\rho,\varphi,z,\omega)+\epsilon_0\,\chi^{ee}_{zz} (\omega,-i\p_{\varphi})\,(\mathbb{D}\mathbf{E})_{z},\nonumber\\
&& P_{x} (\rho,\varphi,z,\omega)=P^{N}_{x} (\rho,\varphi,z,\omega)+\epsilon_0\,\chi^{ee}_{xx}(\omega,-i\p_{\varphi})(\mathbb{D}\mathbf{E})_{x}
+\chi^{ee}_{xy}(\omega,-i\p_{\varphi})(\mathbb{D}\mathbf{E})_{y},\nonumber\\
&& P_{y} (\rho,\varphi,z,\omega)=P^{N}_{y} (\rho,\varphi,z,\omega)+\epsilon_0\,\chi^{ee}_{yx}(\omega,-i\p_{\varphi})(\mathbb{D}\mathbf{E})_{x}
+\chi^{ee}_{yy}(\omega,-i\p_{\varphi})(\mathbb{D}\mathbf{E})_{y},\\
\nonumber
\end{eqnarray}
or in compact form
\begin{equation}\label{matrix}
\mathbf{P} (\mathbf{r},\omega)=\mathbf{P}^{N} (\mathbf{r},\omega)+\epsilon_0\,\boldsymbol{\chi}^{ee} (\omega,-i\p_{\varphi})\cdot\mathbb{D}\mathbf{E}.
\end{equation}
From (\ref{PO-compact},\ref{matrix}) we finally find \cite{Maghrebi}
\begin{equation}\label{FEA}
\Bigl\{\curl\curl\,-\frac{\omega^2}{c^2}\mathbb{I}-\frac{\omega^2}{c^2}\tilde{\mathbb{D}}\cdot \boldsymbol{\chi}^{ee}(\omega,-i \p_\varphi)\cdot\mathbb{D}\Bigl\}\cdot\mathbf{E}=\mu_0\omega^2\tilde{\mathbb{D}}\mathbf{P}^N.
\end{equation}
The presence of operators $\mathbb{D},\tilde{\mathbb{D}}$ in (\ref{FEA}) makes it a complicated equation. For small velocity regime $(v/c\ll 1)$ we can set approximately $\mathbb{D},\tilde{\mathbb{D}}\approx 1 $ and in this case the corresponding dyadic Green's function may be obtained exactly otherwise a perturbative expansion of the dyadic Green's function may be useful \cite{Green}. In high velocity regime numerical calculations may be applied.
%%%%%%%%%%%%%%%%%%%%%%%%%%%%%%%%%%%%%%%%%%%%%%%%%%%%%%%%%%%%%%%%%%%
\section{Fluctuation-dissipation relations}
%%%%%%%%%%%%%%%%%%%%%%%%%%%%%%%%%%%%%%%%%%%%%%%%%%%%%%%%%%%%%%%%%%%
From the definition of polarization and (\ref{HOMO}) we have
\begin{eqnarray}\label{FLUC}
P^{N}_{i,m} (\rho,z,t) &=& \epsilon_0\intf d\nu\,f_{ij}(\nu,t)\,X^{N}_{j,m} (\rho,z,t,\nu)\nonumber\\
&=& \epsilon_0\intf d\nu\,f_{ij}(\nu,t)\,\bigg[a^{\dag}_{j,m}(\rho,z,\nu)e^{i(\nu-m\omega_0)t}+a_{j,-m}(\rho,z,\nu)\,e^{-i(\nu+m\omega_0)t}\bigg],\\
\nonumber
\end{eqnarray}
with Fourier transform
\begin{equation}\label{FFLUC}
P^{N}_{i,m} (\rho,z,\omega)=2\pi\epsilon_0 f_{ij} (m\omega_0-\omega)\,a^{\dag}_{j,m} (\rho,z,m\omega_0-\omega)+2\pi\epsilon_0 f_{ij} (\omega-m\omega_0)\,
a_{j,m} (\rho,z,\omega-m\omega_0).
\end{equation}
If the dielectric is held in temperature $T$, then from (\ref{FLUC-a}, \ref{FFLUC}) we find the following fluctuation-dissipation relations
\begin{eqnarray}\label{FLUC-DISS}
\langle P^{N}_{z}(\mathbf{r},\omega)P^{N\dag }_{z}(\mathbf{r}',\omega')\rangle &=& \hbar\epsilon_0\sum_m e^{i m (\varphi-\varphi')}\Gamma_{zz}(\omega,m)\,\delta(\omega-\omega')\,\frac{\delta(\rho-\rho')\delta(z-z')}{\rho},\nonumber\\
\la P^{N}_{x}(\mathbf{r},\omega)P^{N\dag}_{x}(\mathbf{r}',\omega')\ra &=& \la P^{N}_{y}(\mathbf{r},\omega)P^{N\dag}_{y}(\mathbf{r}',\omega')\ra\nonumber\\
&=& \hbar\epsilon_0\sum_{m} e^{i m (\varphi-\varphi')}\Gamma_{xx}(\omega,m)\,\delta(\omega-\omega')\,\frac{\delta(\rho-\rho')\delta(z-z')}{\rho},\nonumber\\
\la P^{N}_{x}(\mathbf{r},\omega)P^{N\dag}_{y}(\mathbf{r}',\omega')\ra &=& -\la P^{N}_{y}(\mathbf{r},\omega)P^{N\dag}_{x}(\mathbf{r}',\omega')\ra\nonumber\\
&=& \hbar\epsilon_0\sum_{m} e^{i m (\varphi-\varphi')}\Gamma_{xy}(\omega,m)\,\delta(\omega-\omega')\,\frac{\delta(\rho-\rho')\delta(z-z')}{\rho},\nonumber\\
\end{eqnarray}
where $\Gamma_{ij}$ is defined by
\begin{eqnarray}\label{Gama}
\Gamma_{zz} (\omega,m) &=& 2\mbox{Im}[\chi^{0}_{zz} (m\omega_0-\omega)] a_T (m\omega_0-\omega),\nonumber\\
\Gamma_{xx} (\omega,m) &=& \Gamma_{yy} (\omega,m)\nonumber\\
&=&\mbox{Im}[\chi^0_{xx} (m\omega_0-\omega_+)] a_{T} (m\omega_0-\omega_+)+\mbox{Im}[\chi^0_{xx} (m\omega_0-\omega_-)]a_{T} (m\omega_0-\omega_-),\nonumber\\
\Gamma_{xy} (\omega,m) &=& i\mbox{Im}[\chi^0_{xx} (m\omega_0-\omega_-)] a_{T} (m\omega_0-\omega_-)-\mbox{Im}[\chi^0_{xx} (m\omega_0-\omega_+)]a_{T} (m\omega_0-\omega_+),\nonumber\\
\Gamma_{xz} &=& \Gamma_{zx}=\Gamma_{yz}=\Gamma_{zy}=0,
\end{eqnarray}
and $ a_T (\omega)=\coth(\hbar\omega/2k_B T)=2[n_T (\omega)+\frac{1}{2}]$. Using (33) one can rewrite (32) as
\begin{equation}\label{com}
   \la P^{N}_{i}(\mathbf{r},\omega)P^{N\dag}_{j}(\mathbf{r}',\omega')\ra=4\pi\epsilon_0 \hbar\,\Gamma_{ij} (\omega,-i\p_{\varphi})\,\delta(\mathbf{r}-\mathbf{r}')\delta(\omega-\omega').
\end{equation}
%%%%%%%%%%%%%%%%%%%%%%%%%%%%%%%%%%%%%%%%%%%%%%%%%%%%%%%%%%%%%%%%%%%
\section{The Hamiltonian}
%%%%%%%%%%%%%%%%%%%%%%%%%%%%%%%%%%%%%%%%%%%%%%%%%%%%%%%%%%%%%%%%%%%
From canonical conjugate momenta defined in (\ref{Pi},\ref{Q}) we find the following Hamiltonian for the total system
\begin{eqnarray}\label{H}
H &=& \int_{V}  d\mathbf{r}\bigg\{\frac{1}{2\epsilon_0} (\mathbf{P}-\mathbf{D})^2+\frac{1}{2\mu_0} (\curl\mathbf{A})^2+\haf\intf\,d\nu\,[\mathbf{Q}_\nu^2+\nu^2\mathbf{X}_\nu^2]\nonumber\\
&-& \omega_0\intf\,d\nu\,\mathbf{Q}_\nu\cdot\p_{\varphi}\mathbf{X}_\nu -\mathbf{P}\cdot (\mathbf{v}\times\curl\mathbf{A})\bigg\},
\end{eqnarray}
where ${V}$ is the volume of the dielectric. In Hamiltonian (\ref{H}) the first three terms are describing the electromagnetic field quantization in the presence of a non rotating dielectric i.e., $\omega_0=0$ and the last two terms are the modifications caused from rotation. The total interaction between electromagnetic field and the rotating body is
\begin{eqnarray}\label{Int}
 H_{int} &=& -\int_{V_s} d\mathbf{r}[\mathbf{P}(\mathbf{r},t)\cdot\mathbf{E}(\mathbf{r},t)+\mathbf{P}(\mathbf{r},t)\cdot (\mathbf{v}\times\curl\mathbf{A}(\mathbf{r},t))],\nonumber\\
 &=& -\int_{V} d\mathbf{r}[\mathbf{P}(\mathbf{r},t)\cdot (\mathbf{E}(\mathbf{r},t)+\mathbf{v}\times\mathbf{B}(\mathbf{r},t))].\\
 \nonumber
\end{eqnarray}
The rate of work done on the differential volume $d\mathbf{r}$ in the dielectric by the electromagnetic field is $\mathbf{j}\cdot (\mathbf{E}+\mathbf{v}\times\mathbf{B})\, d\mathbf{r}$, where $\mathbf{j}$ is the current density in matter which from (\ref{EA}) is
$\mathbf{j}=\pt\mathbf{P}-\curl (\mathbf{v}\times\mathbf{P})$, therefore the power can be written as
\begin{equation}\label{Power}
\la\mathcal{P}\ra=-\int_{V} d\mathbf{r}\la(\pt\mathbf{P}-\curl (\mathbf{v}\times\mathbf{P}))\cdot (\mathbf{E}+\mathbf{v}\times\mathbf{B})\ra,
\end{equation}
where $|\ra=|\emph{vacuum}\ra_{T_0}\otimes|\emph{matter}\ra_{T}$ is the tensor product of initial thermal states of the electromagnetic and matter fields which are supposed to be held at temperatures $T_0$ and $T$ respectively. The expression for radiated power can be simplified for small bodies or small velocities where we can ignore terms containing the velocity $\mathbf{v}$ and rewrite (\ref{Power}) as
\begin{equation}\label{e38}
\la\mathcal{P}\ra=-\int_{V_s} d\mathbf{r}\la\pt\mathbf{P}\cdot\mathbf{E}\ra.
\end{equation}
For point like particles we have $\mathbb{D},\,\tilde{\mathbb{D}}\approx 1$ and from (\ref{matrix},\ref{FEA}) we find
\begin{eqnarray}\label{e39}
E_i (\mathbf{r},\omega) &=& E_{0,i} (\mathbf{r},\omega)+\mu_0\omega^2\int d\mathbf{r}'\,G_{0,ij} (\mathbf{r},\mathbf{r}',\omega)\,P^{N}_{j} (\mathbf{r}',\omega),\nonumber\\
P_i (\mathbf{r},\omega) &=& P^N_i (\mathbf{r},\omega)+\epsilon_0 \chi^{ee}_{ij} (\omega,-i\p_{\varphi}) E^{N}_{j} (\mathbf{r}',\omega).\\
\nonumber
\end{eqnarray}
If we take the inverse Fourier transform of the second parts of the equations (\ref{e39}) and denote them respectively by $\mathbf{E}^{ind}$ and $\mathbf{P}^{ind}$ as in \cite{Manjavacas}, then
\begin{equation}\label{e40}
\la\mathcal{P}\ra=-\int_{V} d\mathbf{r}\la \pt\mathbf{P}^N \cdot\mathbf{E}^{ind}+\pt\mathbf{P}^{ind}\cdot\mathbf{E}_0 \ra.
\end{equation}
This recent relation applied in \cite{Manjavacas} to find the radiated power from a rotating spherical particle and then the authors reobtained the same results from quantum mechanical considerations. For an extended object, the situation is much more complicated. In this case we need the dyadic Green's function and also we should keep all terms in (\ref{Power}). One can follow a perturbative approach based on the interaction term given in (\ref{Int}) and Dyson-Schwinger formula for Green's function expansion from which for example the radiative process can be determined using Fermi's golden rule.
For an extended body with azimuthal symmetry and small angular velocity from (\ref{e38}) we have
\begin{equation}\label{power1}
  \la\mathcal{P}\ra=\int_{V} d\mathbf{r}\int\int_{-\infty}^{\infty}\frac{d\omega}{2\pi}\frac{d\omega'}{2\pi}e^{-i(\omega+\omega')t}
  (i\omega)\la \mathbf{P}(\mathbf{r},\omega)\cdot\mathbf{E}(\mathbf{r},\omega')\ra,
\end{equation}
using (\ref{matrix},\ref{FEA}) we find
\begin{eqnarray}\label{power2}
  \la\mathcal{P}\ra &=& \int_{V} d\mathbf{r}\int\int_{-\infty}^{\infty}\frac{d\omega}{2\pi}\frac{d\omega'}{2\pi}e^{-i(\omega+\omega')t}
  (i\omega)\bigg[\mu_0 \omega'^{\,2}\int_{V}d\mathbf{r}'\nonumber\\
  && G_{ij}(\mathbf{r},\mathbf{r}',\omega')\la P_i (\mathbf{r},\omega)
  P_j (\mathbf{r}',\omega')\ra+\epsilon_0\chi^{ee}_{ij} (\omega,-i\p_\varphi)
  \la E_j (\mathbf{r},\omega)E_i (\mathbf{r}',\omega')\ra|_{\mathbf{r}'\rightarrow\mathbf{r}}\bigg].\nonumber\\
\end{eqnarray}
Now using (34) and
\begin{equation}\label{dyad}
\la E_j (\mathbf{r},\omega)E_i (\mathbf{r}',\omega')\ra=\frac{2\hbar\omega^2}{\epsilon_0 c^2} \mbox{Im}G_{ji} (\mathbf{r},\mathbf{r}',\omega)
\,\delta(\omega+\omega')\,a_{T_0} (\omega),
\end{equation}
we find the following general formula in terms of the dyadic Green's tensor
\begin{eqnarray}\label{power3}
  \la\mathcal{P}\ra &=& \frac{\hbar}{2\pi c^2}\int_{V} d\mathbf{r}\int_{0}^{\infty} d\omega\,\omega^3
   [a_{T} (\omega-\omega_0 \hat{l}_z)-a_{T_0}(\omega)]\bigg[2\mbox{Im}\chi^{0}_{zz}(\omega-\omega_0 \hat{l}_z)
   \mbox{Im}\,G_{zz}(\mathbf{r},\mathbf{r}',\omega)\nonumber\\
   &+&\mbox{Im}\chi^{0}_{xx}(\omega-\omega_0 \hat{l}_z)[\mbox{Im}\,G_{xx}(\mathbf{r},\mathbf{r}',\omega)
   +\mbox{Im}\,G_{yy}(\mathbf{r},\mathbf{r}',\omega)]\cos(\varphi-\varphi')|\bigg ]_{\mathbf{r}'\rightarrow\mathbf{r}},\nonumber\\
\end{eqnarray}
where $\hat{l}_z=-i\p_{\varphi}$, and we have used the symmetry properties of
tensors $G_{ij} (\mathbf{r},\mathbf{r}',\omega)$ and $\Gamma_{ij} (\omega,-i\p_{\varphi})$. For a small rotating body using $\mbox{Im}\,\alpha_{ij}(\omega)=V\,\mbox{Im}\,\chi^{0}_{ij}(\omega)$, (\ref{power3}) can be rewritten as
\begin{eqnarray}\label{power4}
&& \la\mathcal{P}\ra =\frac{\hbar}{2\pi c^2}\int_{0}^{\infty} d\omega\,\omega^3\bigg[ \big [2\mbox{Im}\alpha_{zz}(\omega)
   \mbox{Im}\,G_{zz}(\omega)\big] [a_{T} (\omega)-a_{T_0}(\omega)]\nonumber\\
  && +[\mbox{Im}\,G_{xx}(\omega) +\mbox{Im}\,G_{yy}(\omega)]\big[\mbox{Im}\alpha_{xx}(\omega_-) [a_{T} (\omega_-)-a_{T_0}(\omega)]+\mbox{Im}\alpha_{xx}(\omega_+) [a_{T} (\omega_+)-a_{T_0}(\omega)]\big]\bigg],\nonumber\\
\end{eqnarray}
where by $G_{ii}(\omega)$ we mean $G_{ii}(\mathbf{r}_0,\mathbf{r}_0,\omega)$ and $\mathbf{r}_0$ is the position of the point-like spinning body.
\theoremstyle{definition}
\newtheorem{exmp}{Example}
\begin{exmp}
As an example let us consider a small rotating dielectric with the susceptibility $\chi(\omega)$ and angular velocity $\omega_0$ in the vicinity of a semi-infinite dielectric with dielectric function
\begin{equation}\label{e}
\varepsilon(\mathbf{r},\omega) = \left\{
\begin{array}{cc}
\varepsilon(\omega) \qquad\quad  z\leqslant 0 \\
1 \qquad\qquad  z> 0,
\end{array}
\right.
\end{equation}
see Fig.1.
\begin{figure}
   \includegraphics[scale=0.4]{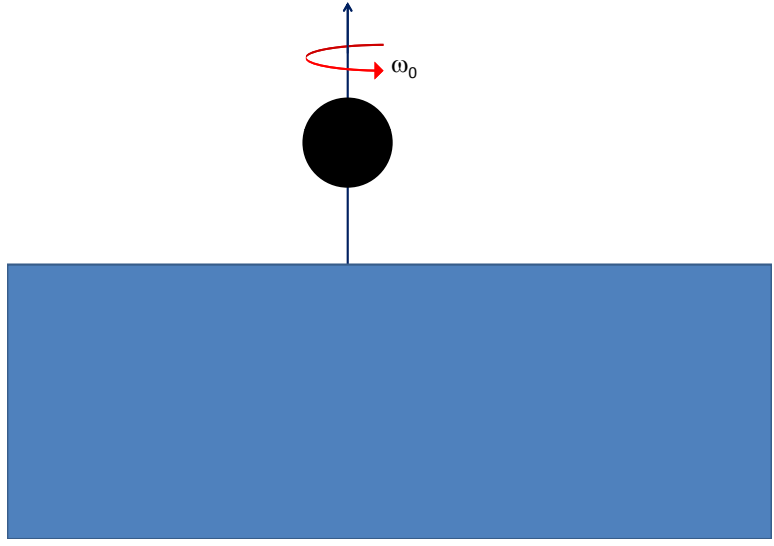}\\
  \caption{(Color online) A rotating sphere in the vicinity of a semi-infinite homogeneous and isotropic matter }\label{setup}
\end{figure}
A related problem with the same geometry has been investigated in \cite{Huth}. The diadic Green's function for this geometry has been calculated in detail in \cite{maradudin}. The necessary components are given by
\begin{eqnarray}\label{green}
G_{xx}(\mathbf{r},\mathbf{r}',\omega)&=&\int_0^\infty\frac{d^2k_\Vert}{(2\pi)^2}e^{i\mathbf{k_\Vert}.
(\mathbf{r}_\Vert-\mathbf{r}_\Vert')}\big [ \frac{k_x^2}{k_\Vert ^2}\,g_{xx}(\mathbf{k}_\Vert, \omega\vert z,z')+\frac{k_y^2}{k_\Vert ^2}\,g_{yy}(\mathbf{k}_\Vert, \omega\vert z,z')\big ] \\ \nonumber
G_{yy}(\mathbf{r},\mathbf{r}',\omega)&=&\int_0^\infty\frac{d^2k_\Vert}{(2\pi)^2}e^{i\mathbf{k_\Vert}.
(\mathbf{r}_\Vert-\mathbf{r}_\Vert')}\big [ \frac{k_y^2}{k_\Vert ^2}\,g_{xx}(\mathbf{k}_\Vert, \omega\vert z,z')+\frac{k_x^2}{k_\Vert ^2}\,g_{yy}(\mathbf{k}_\Vert, \omega\vert z,z')\big ] \\ \nonumber
G_{zz}(\mathbf{r},\mathbf{r}',\omega)&=&\int_0^\infty\frac{d^2k_\Vert}{(2\pi)^2}e^{i\mathbf{k_\Vert}.
(\mathbf{r}_\Vert-\mathbf{r}_\Vert')}\,g_{zz}(\mathbf{k}_\Vert, \omega\vert z,z')
\end{eqnarray}
where
\begin{eqnarray}\label{green1}
g_{xx}(\mathbf{k}_\Vert, \omega\vert z,z')&=&-\frac{2\pi i k c^2}{\omega^2}\big [\frac{k_1+\varepsilon(\omega)k}{k_1-\varepsilon(\omega)k}\, e^{ik(z+z')}+e^{ik\vert z-z'\vert} \big ] \\ \nonumber
g_{yy}(\mathbf{k}_\Vert, \omega\vert z,z')&=&\frac{2\pi i}{k}\big[\frac{k_1+k}{k_1-k}\,e^{ik(z+z')}-e^{ik\vert z-z'\vert}\big]
\\ \nonumber
g_{zz}(\mathbf{k}_\Vert, \omega\vert z,z')&=&\frac{2\pi i k_\Vert^2 c^2}{k\omega^2}\big [\frac{k_1+\varepsilon(\omega)k}{k_1-\varepsilon(\omega)k}\, e^{ik(z+z')}-e^{ik\vert z-z'\vert} \big ]+\frac{4\pi c^2}{\omega^2}\delta(z-z')
\end{eqnarray}
and $k=(\frac{\omega^2}{c^2}-k_\Vert^2)^{\frac{1}{2}}$,  $k_1=-(\frac{\varepsilon(\omega)\omega^2}{c^2}-k_\Vert^2)^{\frac{1}{2}}$.
An interesting limiting case is when $\varepsilon(\omega)\rightarrow \infty$ for $z<0$, i.e., the semi-infinite space is an ideal conductor. In this case, we can easily find the imaginary parts of $g_{xx}(\mathbf{k}_\Vert, \omega\vert z,z')$, $g_{yy}(\mathbf{k}_\Vert, \omega\vert z,z')$ and $g_{zz}(\mathbf{k}_\Vert, \omega\vert z,z')$ for $\mathbf{r}\rightarrow\mathbf{r}'$, as
\begin{eqnarray}\label{green2}
\mbox{Im}\,g_{xx}(\mathbf{k}_\Vert, \omega\vert z,z)&=&-\frac{2\pi k c^2}{\omega^2}[1-\cos(2kz)]  \\ \nonumber
\mbox{Im}\,g_{yy}(\mathbf{k}_\Vert, \omega\vert z,z)&=&-\frac{2\pi}{k}[1-\cos(2kz)] \\ \nonumber
\mbox{Im}\,g_{zz}(\mathbf{k}_\Vert, \omega\vert z,z)&=&-\frac{2\pi k_\Vert^2c^2}{k\omega^2}[1+\cos(2kz)]
\end{eqnarray}
The dielectric function of metals can be described approximately by Drude model in term of the dc conductivity $\sigma_0$ as
\begin{equation}\label{e10}
\epsilon(\omega)\approx\frac{4\pi\sigma_0}{\omega},
\end{equation}
and using the small radius expansion of the Mie scattering coefficient $(a/\lambda\ll 1)$, we find
\begin{equation}\label{e11}
\alpha(\omega)\approx a^3\frac{\epsilon(\omega)-1}{\epsilon(\omega)+2},
\end{equation}
where $a$ is the radius of the rotating sphere. In Fig.2, the radiated power is depicted in terms of the distance from the surface of an ideal metal for a gold nano-particle which tends to the result reported in \cite{Manjavacas} in $z\rightarrow\infty$ as expected.
\begin{figure}
   \includegraphics[scale=0.5]{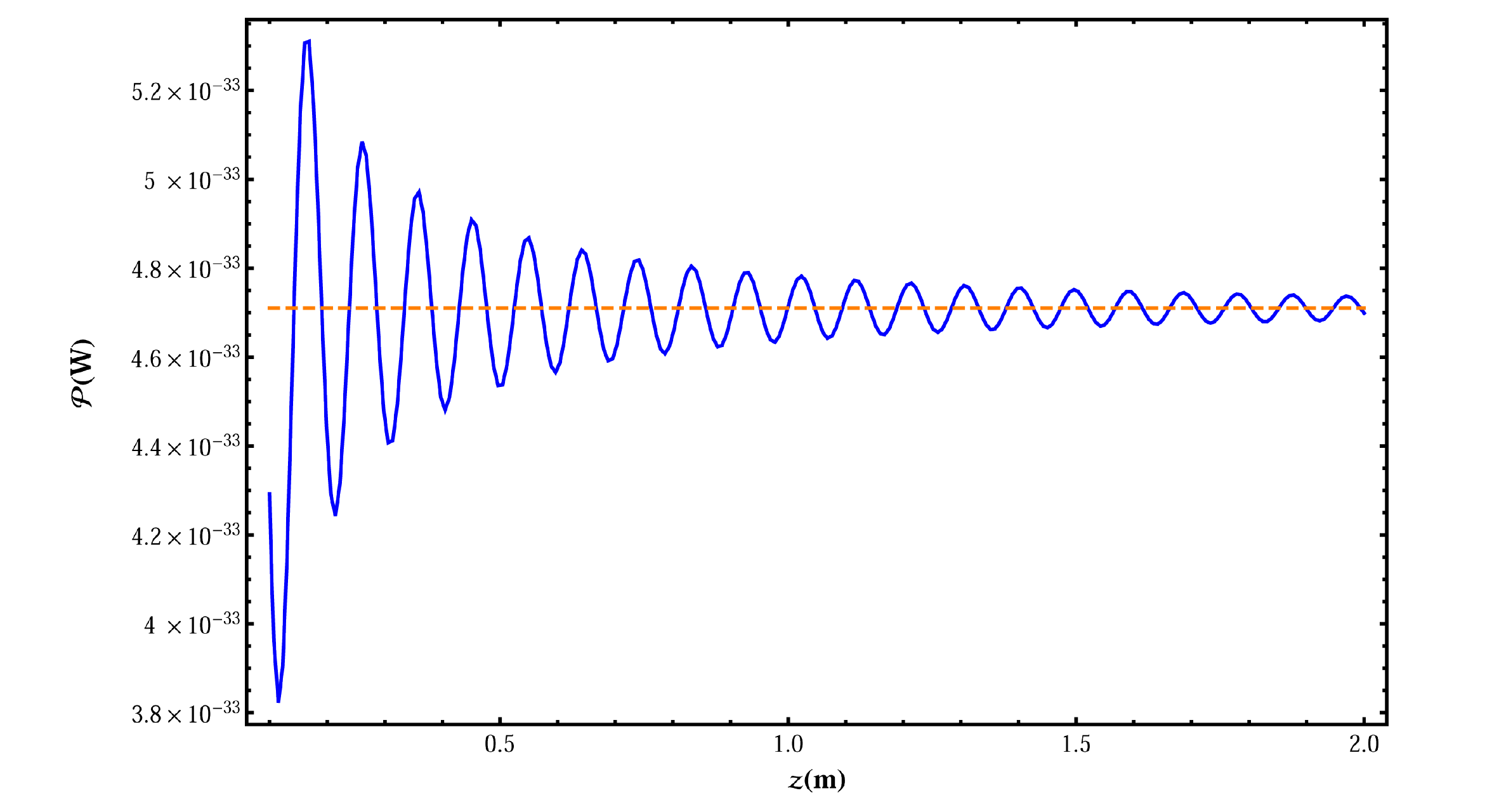}\\
  \caption{ (Color online) The radiation power for a gold nano-particle with $a=10 nm$ and $(\sigma_0\approx1.6\times 10^7 \Omega^{_1}m^{-1})$. The temperatures of the particle and its medium are assumed to be 10 Kelvin and 1 kelvin respectively. The solid line is the radiated power in the presence of a semi-infinite space and the dashed line in vacuum i.e., the result reported in \cite{Manjavacas}. }\label{radi}
  \end{figure}
We can also plot the power spectrum of the rotating particle showing the probability of emitting photons at a specific frequency $\omega$ and compare the results for the case of semi-infinite space and vacuum. Power spectrum contains thermal photons and photons created from rotational motion. In Fig.3, the power spectrums are compared when the thermal photons are extremely more than the photons created from rotational motion of the sphere. In Fig.4, conditions are taken to reduce the thermal photons and increase the photons created from rotational motion.
\end{exmp}
\section{Frictional torque}
The torque produced by an electric field $\mathbf{E}$ on a dipole $\mathbf{P}$ is given by $\mathbf{P}\times\mathbf{E}$. The torque experienced by the particle along the rotation axis $\mathbf{\hat z}$ is
  \begin{equation}
  \mathbf{M}=\int_V d\mathbf{r} \la \mathbf{p}(t)\times \mathbf{E}(\mathbf{r_0},t) \ra\cdot\mathbf{\hat z}.
  \end{equation}
By separating the two contributions we can rewrite this recent relation as
  \begin{eqnarray}
  \mathbf{M}&=&\int_V d\mathbf{r}  \la \mathbf{p}^{N}(t)\times \mathbf{E}^{ind}(\mathbf{r_0},t)+\mathbf{p}^{ind}(t)\times \mathbf{E}^N(\mathbf{r_0},t) \ra . \mathbf{\hat z}\\ \nonumber
&=&\mathbf{M}_P+\mathbf{M}_E
  \end{eqnarray}
The first term accounts for the fluctuations of the particle dipole moment that correlates with the resulting induced field, while the second one involves field fluctuations and the dipole that they have induced. therefore,
  \begin{eqnarray}
  \mathbf{M}_P = \int_V d\mathbf{r} \la \mathbf{p}^{N}(t)\times \mathbf{E}^{ind}(\mathbf{r_0},t) \ra,
  \end{eqnarray}
and using (\ref{e39}) we find
   \begin{eqnarray}\label{torq}
  \mathbf{M}_P &=&\frac{i\mu_0}{8\pi^2} \int_V d\mathbf{r}\int d\mathbf{r}' \int_{-\infty}^{+\infty} d\omega\int_{-\infty}^{+\infty} d\omega' \omega^2 e^{-i(\omega+\omega')}\bigg [ G_{yy}(\mathbf{r},\mathbf{r}',\omega)
 \la \mathbf{P}^N_x(\mathbf{r},\omega) \mathbf{P}^N_y(\mathbf{r}',\omega')\ra\nonumber\\
&-& G_{yz}(\mathbf{r},\mathbf{r}',\omega) \la \mathbf{P}^N_x(\mathbf{r},\omega) \mathbf{P}^N_z(\mathbf{r}',\omega')\ra - G_{xx}(\mathbf{r},\mathbf{r}',\omega') \la \mathbf{P}^N_y(\mathbf{r},\omega) \mathbf{P}^N_x(\mathbf{r}',\omega')\ra\nonumber\\
&-& G_{xz}(\mathbf{r},\mathbf{r}',\omega) \la \mathbf{P}^N_y(\mathbf{r},\omega) \mathbf{P}^N_z(\mathbf{r}',\omega')\ra\bigg].
   \end{eqnarray}
Now using (3,30), the dipole densities in laboratory-frame $ \mathbf{P}$ can be written in terms of the dipole densities in the body frame $ \mathbf{P}' $. For example for the x-component we find
   \begin{eqnarray}
P^{N}_{x,m} (\rho,z,t) &=& \epsilon_0\intf d\nu\,\big [f_{xx}(\nu,t) \cos (\omega_0t)\,X^{N}_{x,m} (\rho,z,t,\nu)+f_{xx}(\nu,t) \sin (\omega_0t)\,X^{N}_{y,m} (\rho,z,t,\nu)\big ]\nonumber\\
&=&\frac{\epsilon_0}{2} \intf d\nu\bigg[\,f_{xx}(\nu,t)\,\big[(a^{\dag}_{x,m}(\rho,z,\nu)e^{i(\nu-m\omega_0+\omega_0)t}+a^{\dag}_{x,m}(\rho,z,\nu)e^{i(\nu-m\omega_0-\omega_0)t}\nonumber\\  &+&a_{x,-m}(\rho,z,\nu)\,e^{-i(\nu+m\omega_0-\omega_0)t}+ a_{x,-m}(\rho,z,\nu)\,e^{-i(\nu+m\omega_0+\omega_0)t})\big]\nonumber\\  &-&\,if_{xx}(\nu,t)\,\big[(a^{\dag}_{y,m}(\rho,z,\nu)e^{i(\nu-m\omega_0+\omega_0)t}-a^{\dag}_{y,m}(\rho,z,\nu)e^{i(\nu-m\omega_0-\omega_0)t}\nonumber\\
&+& a_{y,-m}(\rho,z,\nu)\,e^{-i(\nu+m\omega_0-\omega_0)t}-a_{y,-m}(\rho,z,\nu)\,e^{-i(\nu+m\omega_0+\omega_0)t})\big]\bigg].
   \end{eqnarray}
After some simple algebra and using Fourier transform, it can be easily shown that the dipole moment components in Lab and body frames are related as
  \begin{eqnarray}
  \mathbf{P}_x(\omega) &=& \frac{1}{2}[\mathbf{P}'_x(\omega_+)+i\mathbf{P}'_y(\omega_+)+\mathbf{P}'_x(\omega_-)-i\mathbf{P}'_y(\omega_-)]\\
  \mathbf{P}_y(\omega) &=& \frac{1}{2}[-i\mathbf{P}'_x(\omega_+)+\mathbf{P}'_y(\omega_+)+i\mathbf{P}'_x(\omega_-)+\mathbf{P}'_y(\omega_-)] \\
   \mathbf{P}_z(\omega)&=&\mathbf{P}'_z(\omega).
  \end{eqnarray}
%%%%%%%%%%%%%%%%%%%%%%%%%%%%%%%%%%%%%%%%%%%%%%%%%%%%%%%%%%%%%%%%%%%%%%%%%%%%%%%%%%%%%%%%%%%%%%%%%%%%%%%%%%%%%%%%%%%%%%%%%%%%%%%%%%%%%%%%%%%%%%%%%%%%%%%%%%%%%%
\begin{figure}
    \includegraphics[scale=0.4]{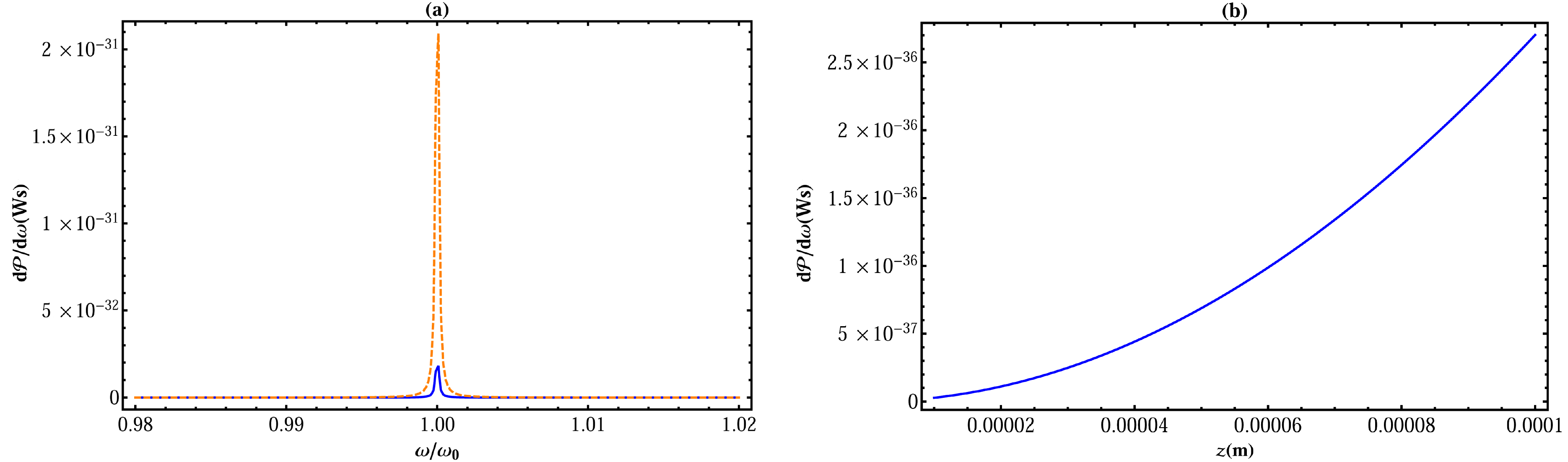}\\
  \caption{(Color online) The power spectrum $\frac{d\la\mathcal{P}\ra}{d\omega}$ radiated by a spinning particle at $T=0$ and $T_0=0.1 K$, where $\theta=\frac{K_BT}{\hbar}$. The solid lines in both figures (a) and (b) represent the power spectrum in the presence of semi-infinite medium and the dashed lines represent the power spectrum in empty space. As we saw in figure(2), the radiation power of the spinning particle in a semi-infinite space oscillates around its value in vacuum as $z$ increases. Figs.3.(a,b) are obtained by setting $z=0.01$ and $z=0.1$, respectively }\label{spectrum}
\end{figure}
%%%%%%%%%%%%%%%%%%%%%%%%%%%%%%%%%%%%%%%%%%%%%%%%%%%%%%%%%%%%%%%%%%%%%%%%%%%%%%%%%%%%%%%%%%%%%%%%%%%%%%%%%%%%%%%%%%%%%%%%%%%%%%%%%%%%%%%%%%%%%%%%%%%%%%%%%%%%%%
The expression for rotational torque can be simplified for small rotating bodies
  \begin{eqnarray}\label{torq5}
  \mathbf{M}_P &=&\frac{i\mu_0}{8\pi^2} \int_V d\mathbf{r}\int d\mathbf{r}' \int_{-\infty}^{+\infty} d\omega\int_{-\infty}^{+\infty} d\omega' \omega^2 e^{-i(\omega+\omega')}\bigg [ G_{yy}(\mathbf{r},\mathbf{r}',\omega)
  \big[\la \mathbf{P}^N_x(\mathbf{r},\omega'_+) \mathbf{P}^N_x(\mathbf{r}',\omega_-)\ra\nonumber\\
&-&\la \mathbf{P}^N_x(\mathbf{r},\omega'_-) \mathbf{P}^N_x(\mathbf{r}',\omega_+)\ra\big] - G_{xx}(\mathbf{r},\mathbf{r}',\omega)\big[\la \mathbf{P}^N_y(\mathbf{r},\omega'_-) \mathbf{P}^N_y(\mathbf{r}',\omega_+)\ra\nonumber\\
&-&\la \mathbf{P}^N_y(\mathbf{r},\omega'_+) \mathbf{P}^N_y(\mathbf{r}',\omega_-)\ra\big]\bigg],
   \end{eqnarray}
using (\ref{Gama},\ref{com}) we find
%%%%%%%%%%%%%%%%%%%%%%%%%%%%%%%%%%%%%%%%%%%%%%%%%%%%%%%%%%%%%%%%%%%%%%%%%%%%%%%%%%%%%%%%%%%%%%%%%%%%%%%%%%%%%%%%%%%%%%%%%%%%%%%%%%%%%%%%%%%%%%%%%%%%%%%%%%%%%%
   \begin{figure}
   \includegraphics[scale=0.4]{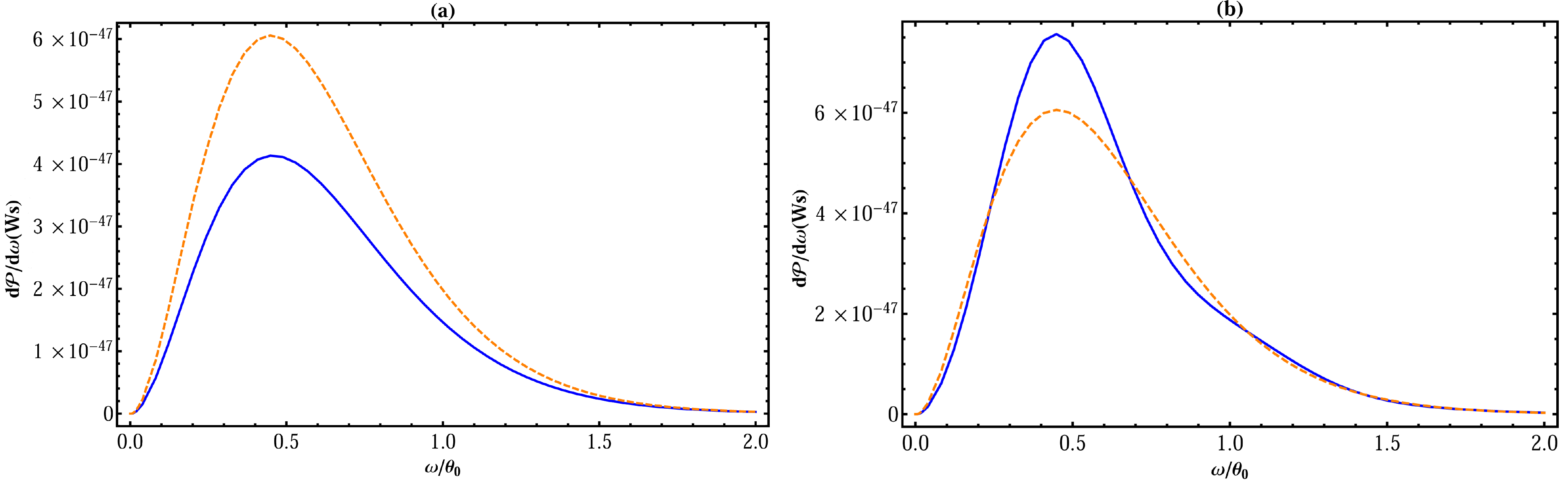}\\
  \caption{(Color online) Figure (a):The power spectrum $\frac{d\la\mathcal{P}\ra}{d\omega}$ radiated by a spinning particle at $T=100$ and $T_0=100 K$. The solid lines represent the power spectrum in the presence of the semi-infinite space and the dashed lines represent the power spectrum in empty space. Here we have set $T=T_0$, to decrease the number of thermal photons. Figure (b) represents the maximum amount of the power spectrum in the presence of the semi-infinite space with respect to the distance from the surface. }\label{spectrum}
  \end{figure}
%%%%%%%%%%%%%%%%%%%%%%%%%%%%%%%%%%%%%%%%%%%%%%%%%%%%%%%%%%%%%%%%%%%%%%%%%%%%%%%%%%%%%%%%%%%%%%%%%%%%%%%%%%%%%%%%%%%%%%%%%%%%%%%%%%%%%%%%%%%%%%%%%%%%%%%%%%%%%
\begin{eqnarray}\label{torq1}
   \mathbf{M}_P =\frac{\hbar}{2\pi c^2}\int_0^\infty d\omega \omega^2 \mbox{Im}[G_{xx}(\omega)+G_{yy}(\omega) ]\,\big[\mbox{Im}\,[\alpha(\omega_+)]a_T(\omega_+)-\mbox{Im}[\alpha(\omega_-)]a_T(\omega_-)\big],
\end{eqnarray}
in a similar way
   \begin{eqnarray}\label{torq2}
   \mathbf{M}_E =\frac{\hbar}{2\pi c^2}\int_0^\infty d\omega \omega^2 \mbox{Im}\,[G_{xx}(\omega) +G_{yy}(\omega) ]\,[\mbox{Im}\,[\alpha(\omega_-)]-\mbox{Im}\,[\alpha(\omega_+) ]a_{T_0}(\omega),
   \end{eqnarray}
therefore,
   \begin{eqnarray}\label{torq3}
    \mathbf{M} =\frac{\hbar}{2\pi c^2}\int_0^\infty d\omega \omega^2 \mbox{Im}\,[G_{xx}(\omega) +G_{yy}(\omega) ]
   \,\bigg[\mbox{Im}\,[\alpha(\omega_+)][a_T(\omega_+)-a_{T_0}(\omega)]\nonumber\\
   -\mbox{Im}\,[\alpha(\omega_-)][a_T(\omega_-)-a_{T_0}(\omega)]\bigg],
   \end{eqnarray}
which coincides with the result obtained in \cite{Manjavacas} for a small spherical body.
%%%%%%%%%%%%%%%%%%%%%%%%%%%%%%%%%%%%%%%%%%%%%%%%%%%%%%%%%%%%%%%%%%%
\section{Conclusions}
%%%%%%%%%%%%%%%%%%%%%%%%%%%%%%%%%%%%%%%%%%%%%%%%%%%%%%%%%%%%%%%%%%%
Starting from a Lagrangian, the electromagnetic field was quantized in the presence of a body rotating along its axis of symmetry. Response functions,  fluctuation-dissipation relations and their connections in body and Lab frames were obtained. A general formula for rotational friction and power radiated by a rotating dielectric body was obtained in terms of the dyadic Green's tensor of the problem. Hamiltonian was determined and possible generalizations were discussed. The case of a small rotating sphere in the vicinity of a semi-infinite space was considered and the rotational friction, radiation power and power spectrum of the rotating sphere were plotted and results compared with those obtained in empty space. The radiation power in semi-infinite space oscillates around the value in vacuum as a result of Interference between the emitted and reflected photons from the surface of the metal.
%%%%%%%%%%%%%%%%%%%%%%%%%%%%%%%%%%%%%%%%%%%%%%%%%%%%%%%%%%%%%%%%%%%
\appendix
\section{Electromagnetic field quantization in a static magnetodielectric medium}
According to Huttner-Barnett model \cite{Huttner}, electromagnetic field quantization in the presence of a linear dielectric and in the absence of external charge and currents can be achieved by considering the total Lagrangian density as
\begin{eqnarray}\label{e1}
 \mathcal{L} &=& \frac{1}{2}\epsilon_0 (-\nabla U-\frac{\partial \mathbf{A}}{\partial t})^2-\frac{1}{2\mu_0} (\nabla\times\mathbf{A})^2\nonumber\\
   &+& \frac{1}{2}\int_0^{\infty} d\nu\,(\rho(\frac{\partial \mathbf{Y}_\nu}{\partial t})^2-\rho\nu^2 (\mathbf{Y}_\nu)^2)\nonumber\\
   &+& \frac{1}{2}\rho(\frac{\partial \mathbf{X}}{\partial t})^2-\frac{1}{2}\rho\nu^2 (\mathbf{X})^2\nonumber\\
   &-& \alpha (\mathbf{A}\cdot\frac{\partial \mathbf{X}}{\partial t} + U\nabla\cdot\mathbf{X})-\int_0^{\infty} d\nu\,\mathbf{X}\cdot\frac{\partial \mathbf{Y}_\nu}{\partial t},
\end{eqnarray}
where the first term is the Lagrangian density of the electromagnetic field, the second therm describes the reservoir or dielectric described by a continuum of harmonic oscillators, the third term is the lagrangian density for the polarization of the medium and the last term describes the interaction between polarization of the medium with electromagnetic field and reservoir respectively. Note that in this model we have to introduce two matter fields to model the polarization and dissipative effects and if we want to include the magnetic properties of the medium we have to introduce two other independent fields which is a generalization of the Huttner-Barnett model introduced in \cite{Kheirandish3}. Equivalently, we can also quantize the electromagnetic field in a linear medium using only two independent fields describing the electric and magnetic properties of the medium. Here we follow the latter approach \cite{Kh-Amooghorban} and for simplicity we assume the medium to be isotropic and homogeneous in region $\Omega$ occupied by matter. The Lagrangian density for the total system in temporal gauge $(A^0=0)$ is given by,
\begin{eqnarray}\label{e2}
  \mathcal{L} &=& \haf\epsilon_0\,(\pt \mathbf{A})^2-\frac{1}{2\mu_0}(\curl\mathbf{A})^2+\haf\intf d\nu \,[(\pt \mathbf{X}_\nu)^2-\nu^2\mathbf{X}_\nu^2]\nonumber\\
  &+& \haf\intf d\nu \,[(\pt \mathbf{Y}_\nu)^2-\nu^2\mathbf{Y}_\nu^2]-\epsilon_0\intf d\nu\,f(\nu,\mathbf{r})\mathbf{X}_\nu\cdot\frac{\p\mathbf{A}}{\p t}\nonumber\\
  &+& \frac{1}{\mu_0}\intf d\nu\,g(\nu,\mathbf{r})\mathbf{Y}_\nu\cdot\curl\mathbf{A}.
\end{eqnarray}
From this Lagrangian density the electric polarization $(\mathbf{P})$ and magnetic polarization $(\mathbf{M})$ densities are defined by
\begin{eqnarray}
  \mathbf{P} &=& \epsilon_0\intf d\nu\,f(\nu,\mathbf{r})\mathbf{X}_\nu\\
  \mathbf{M} &=& \frac{1}{\mu_0}\intf d\nu\,g(\nu,\mathbf{r})\mathbf{Y}_\nu.
\end{eqnarray}
The conjugate momenta are also defined by
\begin{eqnarray}
  \mathbf{\Pi} (\mathbf{r},t) &=& \frac{\p \mathcal{L}}{\p (\pt \mathbf{A})}= \epsilon_0\,(\pt \mathbf{A})-\mathbf{P}=-\mathbf{D}, \\
  \mathbf{Q}_\nu &=& \frac{\p \mathcal{L}}{\p (\pt \mathbf{X}_\nu)}=\pt \mathbf{X}_\nu,\\
  \mathbf{S}_\nu &=& \frac{\p \mathcal{L}}{\p (\pt \mathbf{Y}_\nu)}=\pt \mathbf{Y}_\nu,
\end{eqnarray}
where $\mathbf{D}$ is the displacement field. Now the total system can be quantized by imposing equal-time commutation relations
\begin{eqnarray}
 && [A_i (\mathbf{r},t),\Pi_j (\mathbf{r}',t)]=i\hbar\,\delta_{ij}\,\delta (\mathbf{r}-\mathbf{r}'),\\
 && [X_{\nu,i} (\mathbf{r},t,\nu),Q_{\nu',j} (\mathbf{r}',t,\nu')]=i\hbar\,\delta_{ij}\,\delta (\mathbf{r}-\mathbf{r}')
\delta (\nu-\nu'),\\
&& [Y_{\nu,i} (\mathbf{r},t,\nu),S_{\nu',j} (\mathbf{r}',t,\nu')]=i\hbar\,\delta_{ij}\,\delta (\mathbf{r}-\mathbf{r}')
\delta (\nu-\nu').
\end{eqnarray}
From Lagrangian density (2) we have
\begin{eqnarray}
  \frac{\p \mathcal{L}}{\p \mathbf{A}} &=& -\frac{1}{\mu_0}\curl\curl\mathbf{A}+\curl\mathbf{M}, \\
  \frac{\p \mathcal{L}}{\p \mathbf{X}_\nu} &=& -\nu^2 \mathbf{X}_\nu - \epsilon_0 \,f(\nu,\mathbf{r})\frac{\p\mathbf{A}}{\p t}\\
  \frac{\p \mathcal{L}}{\p \mathbf{Y}_\nu} &=& -\nu^2 \mathbf{X}_\nu+\frac{1}{\mu_0}\,g(\nu,\mathbf{r})\curl\mathbf{A}.
\end{eqnarray}
From equations (5-7) and (11-13) and Euler-Lagrange equations we find the following equations of motion for the electromagnetic and material fields
\begin{eqnarray}
 \frac{1}{c^2}\pt^2 \mathbf{A}+\curl\curl\mathbf{A} &=& \mu_0 (\pt \mathbf{P}+\curl\mathbf{M}), \\
  \pt^2 \mathbf{X}_\nu +\nu^2 \mathbf{X}_\nu &=& -\epsilon_0 \,f(\nu,\mathbf{r})\,\pt\mathbf{A}, \\
  \pt^2 \mathbf{Y}_\nu +\nu^2 \mathbf{Y}_\nu &=& \frac{1}{\mu_0}\, g(\nu,\mathbf{r})\curl\mathbf{A}.
\end{eqnarray}
In frequency space Eqs.(15,16) can be solved easily as
\begin{eqnarray}
  \mathbf{X}_\nu (\mathbf{r},\omega) &=& \mathbf{X}^N_\nu (\mathbf{r},\omega)+\frac{\epsilon_0 f(\nu,\mathbf{r})}{\nu^2-\omega^2}\,\mathbf{E}(\mathbf{r},\omega), \\
  \mathbf{Y}_\nu (\mathbf{r},\omega) &=& \mathbf{Y}^N_\nu (\mathbf{r},\omega)+\frac{1}{\mu_0}\frac{ g(\nu,\mathbf{r})}{\nu^2-\omega^2}\,\mathbf{B}(\mathbf{r},\omega),
\end{eqnarray}
where $\mathbf{X}^N_\nu$ and  $\mathbf{Y}^N_\nu$ are homogeneous or noise solutions in frequency space and from them the noise or fluctuating polarization densities can be determined using Eqs.(3,4). Therefore,
\begin{eqnarray}
  \mathbf{P} (\mathbf{r},\omega) &=& \mathbf{P}^N (\mathbf{r},\omega)+\epsilon_0 ^2\intf d\nu\,\frac{ f^2(\nu,\mathbf{r})}{\nu^2-\omega^2}\,\mathbf{E}(\mathbf{r},\omega), \nonumber\\
   &=&  \mathbf{P}^N (\mathbf{r},\omega)+\epsilon_0 \chi_e (\mathbf{r},\omega)\,\mathbf{E}(\mathbf{r},\omega),\\
  \mathbf{M} (\mathbf{r},\omega) &=& \mathbf{M}^N (\mathbf{r},\omega)+\frac{1}{\mu_0^2}\intf d\nu\,\frac{ g^2(\nu,\mathbf{r})}{\nu^2-\omega^2}\,\mathbf{B}(\mathbf{r},\omega), \nonumber\\
   &=&  \mathbf{M}^N (\mathbf{r},\omega)+\frac{1}{\mu_0} \chi_m (\mathbf{r},\omega)\,\mathbf{B}(\mathbf{r},\omega),
\end{eqnarray}
where
\begin{eqnarray}
  \chi_e (\mathbf{r},\omega) &=& \epsilon_0 \,\intf d\nu\,\frac{ f^2(\nu,\mathbf{r})}{\nu^2-\omega^2-i 0^+}, \\
  \chi_m (\mathbf{r},\omega) &=& \frac{1}{\mu_0}\,\intf d\nu\,\frac{ g^2(\nu,\mathbf{r})}{\nu^2-\omega^2-i 0^+},
\end{eqnarray}
are electric and magnetic susceptibilities respectively and satisfy Kramers-Kronig relations \cite{Kh-Amooghorban}. The electric permittivity and inverse magnetic permeability of the medium are also defined by
\begin{eqnarray}
  \epsilon (\mathbf{r},\omega) &=& 1+\chi_e (\mathbf{r},\omega), \\
  \kappa_m (\mathbf{r},\omega) &=& 1- \chi_m (\mathbf{r},\omega),
\end{eqnarray}
respectively. If we are given definite electric permittivity and inverse magnetic permeability of the medium then we can inverse Eqs.(21,22) and find the corresponding coupling functions as \cite{Kh-Amooghorban}
\begin{eqnarray}
 f(\omega) &=& \sqrt{\frac{2\epsilon_0 \omega}{\pi}\,\mbox{Im}[\chi_e (\omega)]}, \\
 g(\omega) &=& \sqrt{\frac{2\omega}{\pi\mu_0}\,\mbox{Im}[\chi_m (\omega)]}.
\end{eqnarray}
By inserting Eqs.(19,20) into Eq.(14) we find
\begin{equation}\label{main}
 \curl (\kappa_m \curl \mathbf{A})-\frac{\omega^2}{c^2}\,\epsilon\,\mathbf{A}=-i\omega\,\mu_0\,\mathbf{P}^N+\mu_0\,\curl\mathbf{M}^N=\boldsymbol{\xi}^N (\mathbf{r},\omega),
\end{equation}
with the formal solution
\begin{equation}\label{solution}
 A_i (\mathbf{r},\omega)=A^H_i (\mathbf{r},t)+\int_\Omega d \mathbf{r}'\,G_{ij} (\mathbf{r},\mathbf{r}',\omega)\,\xi^N_j (\mathbf{r}',\omega),
\end{equation}
where $\mathbf{A}^H$ is the homogeneous solution and $G_{ij} (\mathbf{r},\mathbf{r}',\omega)$ is the dyadic Green's tensor which satisfies the equation
\begin{equation}\label{Green}
  \big[\curl (\kappa_m \curl .)-\frac{\omega^2}{c^2}\,\epsilon\,\big]\cdot \mathbf{G}=\delta (\mathbf{r}-\mathbf{r}')\,\mathbb{I},
\end{equation}
and fulfils all boundary conditions of the problem. This boundary conditions will be imposed by continuity of tangential and normal components of electric and magnetic fields respectively.
\subsection{Hamiltonian}
From conjugate momenta (5-7) we find the Hamiltonian density as
\begin{eqnarray}
  \mathcal{H} &=& \mathbf{\Pi}\cdot\pt\mathbf{A}+\int_0^\infty d\nu\,\mathbf{Q}_\nu \cdot\pt\mathbf{X}_\nu+\int_0^\infty d\nu\,\mathbf{S}_\nu \cdot\pt\mathbf{Y}_\nu-\mathcal{L},\nonumber\\
   &=& \frac{(\mathbf{\Pi}+\mathbf{P})^2}{2\epsilon_0}+\frac{1}{2\mu_0}\,(\curl\mathbf{A})^2+\frac{1}{2}\int_0^{\infty} d\nu\,\big[\mathbf{Q}^2_\nu+\nu^2 \mathbf{X}^2_\nu\big]\nonumber\\
   &+&  \frac{1}{2}\int_0^{\infty} d\nu\,\big[\mathbf{S}^2_\nu+\nu^2 \mathbf{Y}^2_\nu\big]-\mathbf{M}\cdot\curl\mathbf{A}.
\end{eqnarray}
And equivalently we find the equations of motion for electromagnetic and material fields from Hamiltonian $H=\int d\mathbf{r}\,\mathcal{H}$ as \cite{Kheirandish1}
\begin{eqnarray}
  i\hbar\,\pt\mathbf{A} &=& [\mathbf{A},H]\rightarrow \mathbf{D}=\mathbf{P}+\epsilon_0 \mathbf{E},\\
  i\hbar\,\pt\mathbf{\Pi} &=& [\mathbf{\Pi},H]\rightarrow \curl\mathbf{B}=\mu_0\,\pt\mathbf{D}+\mu_0 \curl\mathbf{M}.\\
  \nonumber
\end{eqnarray}
From Eq.(32) we find $\nabla\cdot\mathbf{D}=0$, (no external charges) and $\mathbf{H}=\frac{1}{\mu_0}\mathbf{B}-\mathbf{M}$, \cite{Kheirandish1}.
\acknowledgments
%%%%%%%%%%%%%%%%%%%%%%%%%%%%%%%%%%%%%%%%%%%%%%%%%%%%%%%%%%%%%%%%%%%
This work was done while the first author was enjoying the hospitality
and partial support of the Department of Physics and
Astronomy of the University of Oklahoma. Thanks go, in
particular, to Prof. Kimball A. Milton and his group who made this
very pleasant and exciting sabbatical leave possible.
%%%%%%%%%%%%%%%%%%%%%%%%%%%%%%%%%%%%%%%%%%%%%%%%%%%%%%%%%%%%%%%%%%%

\end{document}